\documentclass[12pt]{iopart}
\usepackage{graphicx,color}
\usepackage{dsfont}
\usepackage{iopams}
\usepackage{times}
 \usepackage{textcomp}
 \usepackage{rotating}
\usepackage{here}
\usepackage{dcolumn}

\newcommand{\el}{\hat{e}_0^{\phantom \dag}}
\newcommand{\gl}{\hat{g}_0^{\phantom \dag}}
\newcommand{\er}{\hat{e}_1^{\phantom \dag}}
\newcommand{\gr}{\hat{g}_1^{\phantom \dag}}

\newcommand{\eld}{\hat{e}_0^{\dag}}
\newcommand{\gld}{\hat{g}_0^{\dag}}
\newcommand{\erd}{\hat{e}_1^{\dag}}
\newcommand{\grd}{\hat{g}_1^{\dag}}

\newcommand{\bra}[1]{\langle{#1}|}
\newcommand{\ket}[1]{|{#1}\rangle}

\newcommand{\nicefrac}[2]{\leavevmode\kern.1em\raise.5ex\hbox{
    \the\scriptfont0 #1}\kern-.1em/\kern-.15em\lower.25ex\hbox{
    \the\scriptfont0 #2}}

\newcommand{\ie}{\textit{i.e., }}

\newcommand{\YBCO}{YBa$_2$Cu$_3$O$_7$}

\newcommand{\BSCCO}{Bi$_2$Sr$_2$Ca$_1$Cu$_2$O$_{8}$}

\newcommand{\NCCO}{Nd$_{2-x}$Ce$_x$CuO$_4$}

\begin{document}

\title{Semifluxons in Superconductivity and Cold Atomic Gases} \author{
  R.~Walser$^1$, E.~Goldobin$^2$, O. Crasser$^1$, D.~Koelle$^2$,
  R.~Kleiner$^2$, W.~P.~Schleich$^1$} \address{$^1$ Institut f\"ur
  Quantenphysik, Universit\"at Ulm, Albert-Einstein Allee 11, D-89069 Ulm,
  Germany } \address{$^2$ Physikalisches Institut II, Universit\"at
  T\"ubingen, Auf der Morgenstelle 14, D-72076 T{\"u}bingen, Germany }
\ead{Reinhold.Walser@uni-ulm.de}

\begin{abstract}
  Josephson junctions and junction arrays are well studied devices in
  superconductivity. With external magnetic fields one can modulate the phase
  in a long junction and create traveling, solitonic waves of magnetic flux,
  called fluxons.  Today, it is also possible to device two different types of
  junctions: depending on the sign of the critical current density
  $j_c\gtrless 0$, they are called $0$- or $\pi$-junction. In turn, a $0$-$\pi$
  junction is formed by joining two of such junctions.  As a result, one
  obtains a pinned Josephson vortex of fractional magnetic flux, at the
  $0$-$\pi$ boundary. Here, we analyze this arrangement of superconducting
  junctions in the context of an atomic bosonic quantum gas, where two-state
  atoms in a double well trap are coupled in an analogous fashion.  There, an
  all-optical $0$-$\pi$ Josephson junction is created by the phase of a
  complex valued Rabi-frequency and we a derive a discrete four-mode model for
  this situation, which qualitatively resembles a semifluxon.
\end{abstract}

{\today}


\maketitle
\section{Introduction}
During the past two decades, the field of cold atomic gases has come a long
way starting from almost lossless trapping and cooling techniques
\cite{enricofermiproc92} to reaching quantum degeneracy of Bosons and Fermions
\cite{stringaribuch}. Many phenomena that are the hallmarks of condensed
matter physics, whether in superfluid or superconducting materials
\cite{kleinerbuch}, are revisited within this novel context. Due to the
remarkable ease with which it is possible to isolate the key mechanisms from
rogue processes, one can clearly identify phase transitions, for example,
Bose-Einstein condensation, the Mott phase transition or the BEC-BCS
crossover.  Today however, degenerate gases are still at a disadvantage if we
consider robustness, portability or the ability for a mass production compared
to solid-state devices, which is the great achievement of the semiconductor
industry. Strong attempts to miniaturize cold gas experiments
\cite{folman02,fortagh07} and to make them portable
\cite{anderson04,becmugrav06,nandi06,koenemann07} are currently under way in
many laboratories. 

Due to the great importance and practical relevance of the Josephson effect in
superconducting systems \cite{likharev79,barone82,schoen01}, it has also
received immediate attention after the first realization of Bose-Einstein
condensates
\cite{castin97,ketterleinterference,ketterleProc99,Anderson1998a,sols98,williams199,williams499,leggett401,giovanazzi500,oberthaler05,oberthaler07,
  inguscio801,nandi07}.  In particular, the combination of optical lattices
with ultracold gases \cite{bloch05,bloch07} has boosted the possibilities to
investigate junction arrays experimentally. Remarkably, even the absence of
phase-coherence between neighboring sites can lead to interference as
demonstrated in \cite{dalibard04}. The possibility to study atomic Josephson
vortices in the mean field description was raised first in connection with the
sine-Gordon equation \cite{kaurov:013627,kaurov:011601}.

In the present article we will report on such a transfer of concepts from a
superconducting device \cite{Goldobin:SF-Shape}, \ie in various realizations
of Josephson junction arrays and their unusual state properties of traveling
(fluxons) and pinned (semifluxons) magnetic flux quanta to an analogous set up
for neutral bosonic atoms in a trap.  In particular, we will investigate an
all-optical $0$-$\pi$ Josephson junction that can be created with a jumping
phase of an optical laser.

This article organized as follows: in Sec.~\ref{flux}, we will give a brief
review of the current status of the superconductor physics of Josephson
junctions. In particular, we will refer to the most relevant publications in
this thriving field of fluxon and semifluxon physics; in Sec.~\ref{becflux},
we will discuss a similar setup, which allows to find a pinned semifluxon in
an atomic $0$-$\pi$ Josephson junction and we compare the results.  Finally,
we will discuss further open questions in a Conclusion.

\section{Fluxons and Semifluxons in Superconductivity}
\label{flux}
The Josephson effect is a well established phenomenon in the solid state
physics. A Josephson junction (JJ) consists of two weakly coupled
superconducting condensates. JJs are usually fabricated artificially using
low- or high-$T_c$ superconducting electrodes separated by a thin insulating
(tunnel), normal metal, or some other (exotic) barrier. JJs can also be
present intrinsically in a anisotropic layered high-$T_c$ superconductors such
as \BSCCO \cite{Kleiner:1992:IJE,Kleiner:1994:IJE}.

The dc Josephson effect (flow of current through a Josephson junction without
producing voltage drop, i.e. without dissipation) is expressed using the first
Josephson relation, which in the simplest case has the form
\begin{equation}
  I_s = I_c \sin(\phi)
  , \label{Eq:CPR0}
\end{equation}
where $I_s$ is the supercurrent flowing through the junction, $I_c$ is the
critical current, i.e. the maximum supercurrent which can pass through the JJ,
and $\phi=\theta_2-\theta_1$ is the difference between the phases of the
quantum mechanical macroscopic wave functions
$\psi_{1,2}=\sqrt{n_s}e^{i\theta_{1,2}}$ of the superconducting condensates in
the electrodes.

Recent advances in physics and technology allow to fabricate and study the
so-called $\pi$-Josephson junctions --- junctions which formally have negative
critical current $I_c<0$. This can be achieved by using a ferromagnetic
barrier, i.e. in Superconductor-Ferromagnet-Superconductor
 (SFS) \cite{Ryazanov:2001:SFS-PiJJ,Blum:2002:IcOscillations,
Bauer:2004:SFS-SpontSuperCurrents,Sellier:2004:SFS:HalfIntShapiro,
Oboznov:2006:SFS-Ic(dF)} or 
Superconductor-Insulator-Ferromagnet-Superconductor (SIFS)
\cite{Kontos:2002:SIFS-PiJJ,Weides:2006:SIFS-HiJcPiJJ} structures. One can
also achieve the same effect using a barrier which effectively flips the spin
of a tunneling electron, e.g. when the barrier is made of a ferromagnetic
insulator \cite{Vavra:2006:SIfIS}, of a carbon
nanotube \cite{Cleuziou:2006:CNT-SQUID} or of a quantum dot created by gating a
semiconducting nanowire \cite{vanDam:2006:QuDot:SuperCurrRev}.

The change in the sign of a critical current has far going consequences. For
example, analyze the Josephson energy (potential energy related to the
supercurrent flow). In a conventional JJ with $I_c>0$
\begin{equation}
  U(\phi)=E_J(1-\cos\phi)
  \label{Eq:U_0(phi)}
\end{equation}
and has a minimum at $\phi=0+2\pi n$ (the ground state), where
$E_J=\Phi_0I_c/2\pi$ is the Josephson energy. If $I_c<0$, we define
$E_J=\Phi_0|I_c|/2\pi>0$ and
\begin{equation}
  U(\phi)=E_J(1+\cos\phi)
  . \label{Eq:U_pi(phi)}
\end{equation}
Obviously, the minimum of energy is reached for $\phi=\pi+2\pi n$. Thus,
in the ground state (the JJ is not connected to a current source, no current
flows through it) the phase drop across a conventional JJ with $I_c>0$ is
$\phi=0+2\pi n$, while for a junction with $I_c<0$ it is
$\phi=\pi+2\pi n$. Therefore, one speaks about ``$0$-JJs'' and
``$\pi$-JJs''.

Further, connecting the two superconducting electrodes of a $\pi$-JJ by a not
very small inductor $L$ (superconducting wire), the supercurrent $\propto
\pi/L$ will start circulating in the loop. Note, that this supercurrent is
spontaneous, \ie it appears by itself, and has a direction randomly chosen
between clockwise and counterclockwise \cite{Bulaevskii:pi-loop}. The magnetic
flux created by this supercurrent inside the loop is equal to $\Phi_0/2$,
where $\Phi_0=h/2e\approx 2.07\cdot 10^{-15}$ Wb is the magnetic flux quantum. Thus, the $\pi$-JJ works as a
{\em phase battery}. This phase battery will work as described, supplying a
supercurrent through the loop with inductor, provided the inductance
$L\gg\Phi_0/I_c$. If the inductance $L$ is not that large, the battery will be
over loaded, providing a smaller phase drop and supporting smaller current.
For very small inductance the battery will stop working completely.
\begin{figure}[!htb]
  \begin{center}
    \includegraphics[width=0.5\columnwidth]{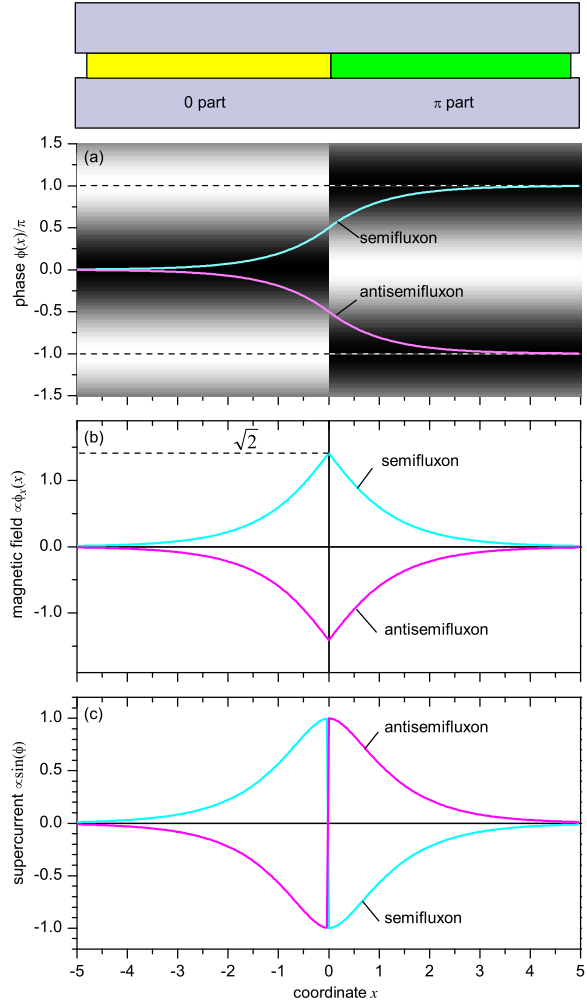}
  \end{center}
  \caption{%
    Sketch of a $0$-$\pi$ JJ and profiles of (a) the phase $\phi(x)$, (b) the
    magnetic field $\propto d\phi(x)/dx$ and (c) the supercurrent density
    $j_s(x)\propto \sin\phi(x)$ corresponding to a semifluxon (black) and
    antisemifluxon (gray).  }
  \label{Fig:Semifluxon}
\end{figure}

Similar effects can be observed in a $\pi$ dc superconducting quantum
interference device (SQUID: one $0$-JJ, one $\pi$-JJ and an inductor $L$
connected in series and closed in a loop) or in $0$-$\pi$ Josephson junction.
Let us focus on a latter case.

Consider a long (along $x$) Josephson junction (LJJ) one half of which at
$x<0$ has the properties of a $0$-JJ (critical current {\em density} $j_c>0$)
and another half at $x>0$ has the properties of a $\pi$-JJ (critical current
{\em density} $j_c<0$). Long means that the length is much larger than the so
called Josephson length $\lambda_J$, which characterizes the size of a
Josephson vortex; typically $\lambda_J\sim10-20 \,\mu$m. What will be the
ground state of such a 0-$\pi$ LJJ?  It turns out that if the junction is long
enough (formally infinitely long), then far away from the $0$-$\pi$ boundary
situated at $x=0$, \ie at $x\to\pm\infty$ the phase $\phi$ will have the
values 0 or $\pm\pi$ (we omit $2\pi n$ here), while in the vicinity of
$0$-$\pi$ boundary the phase $\phi(x)$ smoothly changes from $\phi(-\infty)=0$
to $\phi(+\infty)=\pm\pi$, see Fig.~\ref{Fig:Semifluxon}a.  The exact profile
can be derived analytically
\cite{Bulaevskii:0-pi-LJJ,Xu:SF-shape,Goldobin:SF-Shape}. Since the phase
bends, the local magnetic magnetic field $H\propto d\phi/dx$ will be localized
in the vicinity of the 0-$\pi$ boundary and carry the total flux equal to
$\pm\Phi_0/2$, see Fig.~\ref{Fig:Semifluxon}b. The sign depends on whether the
phase bends from $\phi(-\infty)=0$ to $\phi(+\infty)=+\pi$ or to
$\phi(+\infty)=-\pi$. Thus, such an object is called a semifluxon or an
antisemifluxon. If one analyzes the Josephson supercurrent density flowing
though the barrier $j_s(x)=j_c(x)\sin\phi(x)$, one can see in
Fig.~\ref{Fig:Semifluxon}c that the supercurrent has different directions on
different sides from the $0$-$\pi$ boundary. Since we do not apply any
external current, the flow of current should close in the top and bottom
electrodes, i.e. the supercurrent circulates (counter)clockwise in the case of
(anti)semifluxon.  Thus, {\em a semifluxon is a Josephson vortex of
  supercurrent. It is pinned at the $0$-$\pi$ boundary and has two degenerate
  ground states with the localized magnetic field carrying the flux
  $\pm\Phi_0/2$.}

Semifluxons in various types of JJ has been actively investigated during the
last years. In fact, the first experiments became possible because of deeper
understanding the symmetry of the superconducting order parameter in cuprate
superconductors. This order parameter with the so-called $d$-wave symmetry is
realized in anisotropic superconductors, such as {\YBCO}
or \NCCO. It allowed to
fabricate $0$-$\pi$ grain boundary LJJs
\cite{Kirtley:SF:T-dep,Kirtley:SF:HTSGB} and, later, more controllable
d-wave/s-wave ramp zigzag JJs \cite{Smilde:ZigzagPRL,Ariando:Zigzag:NCCO} and
directly see and manipulate semifluxons using a SQUID microscope
\cite{Hilgenkamp:zigzag:SF,Kirtley:2005:AFM-SF}.

Semifluxons are very interesting non-linear objects: they can form a variety
of ground states
\cite{Kogan:3CrystalVortices,Zenchuk:2003:AnalXover,Susanto:SF-gamma_c,Kirtley:IcH-PiLJJ},
may flip \cite{Hilgenkamp:zigzag:SF,Kirtley:SF:T-dep} emitting a fluxon
\cite{Goldobin:F-SF,Susanto:SF-gamma_c,Lazarides:Ic(H):SF-Gen}, or be
rearranged \cite{Goldobin:SF-ReArrange} by a bias current. Huge arrays of
semifluxons were realized \cite{Hilgenkamp:zigzag:SF} and predicted to behave
as tunable photonic crystals \cite{Susanto:2005:1D-FractVortexCrystal}.
Semifluxons are also promising candidates for storage devices in the classical
or quantum domain and can be used to build qubits
\cite{Goldobin:2005:MQC-2SFs} as they behave like macroscopic spin $1/2$
particles.

Now, an interesting question arises: Can one realize $\pi$ or even
0-$\pi$ JJs in an atomic BEC? In the latter case, the degenerate ground state
corresponding to a semifluxon should have a non-trivial spatial phase profile
and semifluxon physics can also be studied using BEC implementation.

\section{Semifluxons in Bose-Einstein Condensates}
\label{becflux}
Here, we will address this question and examine a configuration were the
two-state atoms are trapped in a quasi-one dimensional cigar-shaped trap with
an additional superimposed double well potential in the longitudinal
direction.  The spatial localization of the two-state atoms inside the double
well potential leads to two internal atomic Josephson junctions that are
driven via an optical, complex valued ``$0$-$\pi$'' laser field and they are
motionally connected via tunneling.

First, we will describe details of the model and introduce the Hamiltonian of
the system.  Then, we will examine the classical limit of the field theory and
study the ground state of the Gross-Pitaevskii equation. Finally, we will
exploit the fact that the spatial wavefunctions are localized inside a deep
double well and study a simple four-mode quantum model derived from a Wannier
basis state representation.

\subsection{$0$-$\pi$-junction in a Bose-Einstein condensate}
\begin{figure}[h]
  \input{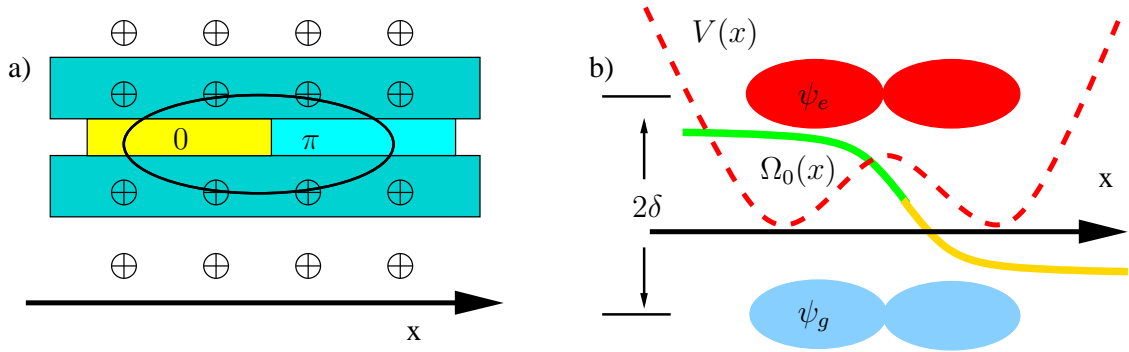}
  \caption[Analogy of superconductor and an atomic BEC]{\label{becmod} 
    Analogy of a $0$-$\pi$ Josephson junction in a superconductor (a) with an
    atomic two-level BEC (b) in a double well trap  $V(x)$ with a position
    dependent Rabi frequency $\Omega_0(x)$.  The excited atomic level
    $\ket{e}$ is separated from the ground level $\ket{g}$ by the detuning
    $2\delta$.}
\end{figure}

To model the $0$-$\pi$-Junction \cite{kaurov:013627,kaurov:011601} in a BEC,
we are guided by the the condensed matter physics setup depicted in
Fig.~\ref{becmod}. As in the single atom case, we replace the two
superconductors by an atomic two-level BEC in a cigar shaped trap. The two
states of the atom, \ie the excited state $\ket{e}$ and ground state $\ket{g}$
couple via a position dependent Rabi frequency $\Omega_0(x)$, which exhibits a
phase jump at the origin of the $x$-axes
\begin{equation}
\label{ome}
  \Omega_0(x)=\left\{\begin{array}{ll}\Omega_0,&x<0,\\
      \Omega_1=-\Omega_0,&x\geq 0.\end{array}\right.
\end{equation}

In this quasi one-dimensional scenario, we will represent the two-state atoms
by a spinorial bosonic quantum field $\hat{\Psi}$, which satisfies the
commutator relation
\begin{equation}
  \label{eq:commut}
[\hat{\Psi}(x),\hat{\Psi}(y)^{\dagger}]= \mathds{1} \delta(x-y).
\end{equation}
This field can be decomposed in any complete single particle basis
$\ket{\sigma,l}$, which resolves the spatial extent of the field and the 
internal structure of the atoms, \ie
\begin{equation}
  \label{eq:psiop}
  \hat{\Psi}=\sum_{\sigma=\{e,g\}}
  \sum_{l=0}^{\infty}
 \ket{\sigma,l}\,   \hat{a}_{\sigma,l}
\end{equation}
and we denote the corresponding discrete bosonic field amplitudes by
$\hat{a}_{\sigma,l}$.  Here, $\sigma$ characterizes the internal states by
$(e,g)$ and the external motion in the double well potential by a quantum
label $l$.  The dynamical evolution of the atomic field is governed by
following Hamiltonian
\begin{eqnarray}
  \label{eq:ham}
  \hat{H}&=&\int_{-\infty}^{\infty} {\rm d}x\, 
  \hat{\Psi}^\dag(x) \left[
    -\partial^2_x+V(x)+
    \left(
      \begin{array}{cc}\delta &\Omega_0(x)\\
        \Omega_0^\ast(x)&-\delta
      \end{array}
    \right)\right]
  \hat{\Psi}(x)\\\nonumber
  &&+g\hat{\Psi}_e^\dagger(x)\hat{\Psi}_e^\dagger(x)
    \hat{\Psi}_e^{\phantom \dagger}(x)\hat{\Psi}_e^{\phantom \dagger}(x)
    +g\hat{\Psi}_g^\dagger(x)\hat{\Psi}_g^\dagger(x)
    \hat{\Psi}_g^{\phantom \dagger}(x)\hat{\Psi}_g^{\phantom \dagger}(x).
\end{eqnarray}
In here, we use dimensionless units, in particular we have set $\hbar=1$ and
the mass of the atom $m=1$.  The energy consists of the single particle energy
in a trap $V(x)$, which is identical for both species, the electric dipole
interaction of the two-state atom \cite{schleich01}, as well as a generic
collision energy proportional to the coupling constant $g=g_{ee}=g_{gg}$. To
simplify the analysis, we have deliberately set the cross-component scattering
length $g_{eg}=0$.  No unaccounted loss channels are present. Therefore, we
have number conservation
\begin{equation}
  \label{numsym}
  \left[\hat{H},\hat{N}\right]=0,
\end{equation}
as a symmetry. If we denote a generic state of the many-particle system in
Fock-space by
\begin{equation}
  \label{eq:fock}
\ket{\psi(t)}=\sum_{\mathbf{n}=0}^\infty
\psi_{\mathbf{n}}(t) \ket{\mathbf{n}=(n_0,n_1,\ldots)},  
\end{equation}
 then one can obtain the dynamics of the system most
generally from the Lagrangian formulation 
\begin{equation}
  \label{eq:lag}
  \mathcal{L}[\psi_\mathbf{n}(t),\dot{\psi}_\mathbf{n}(t)]=
\bra{\psi(t)}i\partial_t-\hat{H}\ket{\psi(t)}.
\end{equation}
In this field theory, the canonical momentum is given by
$\pi_\mathbf{n}=\frac{\delta \mathcal{L}}{\delta \dot{\psi}_\mathbf{n}}= i
\psi^\ast_\mathbf{n}$. From the Hamilton equation $\dot{\pi}_\mathbf{n}= 
-\frac{\delta \mathcal{L}}{\delta \psi_\mathbf{n}}$, one recovers the
conventional Schr\"odinger equation in Fock space
\begin{equation}
  \label{eq:schrodinger}
  i\partial_t\ket{\psi}=\hat{H}\ket{\psi}.
\end{equation}
The Lagrangian approach is obviously a central concept in the path integral
formulation of quantum mechanics \cite{feynman}. However, it is also of great
utility in the approximate description of the dynamics if we connect it with
concepts of classical mechanics as we will see in the following.
\subsection{Spatially extended classical model: the Gross-Pitaevskii equation}
The classical limit of the field equations \cite{stringaribuch} can
be recovered quickly by approximating the state of the system by a coherent
state
\begin{equation}
  \label{eq:coh}
  \hat{\Psi}_\sigma(x)\ket{\psi}=\psi_\sigma(x)\ket{\psi}.  
\end{equation}
Within this approximation, we obtain from the Lagrangian of Eq.~(\ref{eq:lag})
the two component Gross-Pitaevskii equation
$\psi=(\psi_e(x,t),\psi_g(x,t))^\top$
\begin{equation}
  i\partial_t\psi=
\left[
  -\partial^2_x+V(x)+
  \left(
    \begin{array}{cc}\delta +2g|\psi_e|^2&\Omega_0(x)\\
      \Omega^\ast_0(x)&-\delta+2g|\psi_g|^2
    \end{array}
  \right)\right]
\psi.
\end{equation}
For a macroscopically occupied field this equation models the spatial
evolution of the coupled Josephson junctions very well 
\cite{sols98,williams199,williams499, leggett401,giovanazzi500,
  oberthaler05,oberthaler07, inguscio801,kaurov:011601}. 
\subsection{Discrete quantum model: two coupled Josephson junctions}
\begin{figure}[h]
  \begin{center}
    \includegraphics[angle=-90,width=0.5\columnwidth]{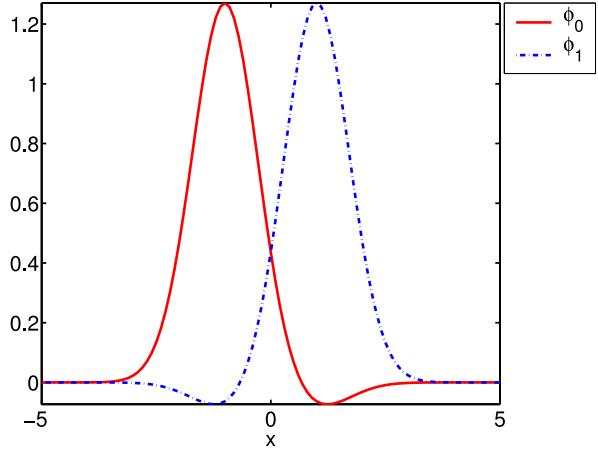}
    \caption[Even and odd mode.]{
      From even and odd parity modes of a double-well potential, one can
      construct the left $\phi_0$ and right $\phi_1$ localized Wannier
      states of the system.}
    \label{wan}
\end{center}
\end{figure}

To gain more insight into the quantum properties of the groundstate of the
system \cite{Goldobin:2005:MQC-2SFs}, one can decompose the field into its
principal components and disregard small corrections.  A double well potential
can be considered as the limiting case of a periodic lattice. While even and
odd parity modes relate to delocalized Bloch-states in a periodic system, left
and right localized modes \ie $\phi_0$ and right $\phi_1$, resemble the
Wannier basis.  These localized modes are depicted in Fig.~\ref{wan}. With
respect to these basis states, we can approximate the field with four modes
\begin{eqnarray}
  \hat{\Psi}_e(x)&=&\el \phi_0(x)+\er \phi_1(x)+\delta\hat{\Psi}_e,\\
  \hat{\Psi}_g(x)&=&\gl \phi_0(x)+\gr \phi_1(x)+\delta\hat{\Psi}_g.
\end{eqnarray}
The four bosonic amplitudes $\{\gl, \el, \gr, \er\}$ satisfy the usual
commutation relations and we will disregard the small corrections of order 
$\delta\hat{\Psi}_\sigma$

If this approximate field is substituted in the Hamiltonian,
Eq.~(\ref{eq:ham}), we can exploit the orthogonality of the wave functions and
obtain the two-body matrix elements $\phi_{ijkl}$
\begin{equation}
  \delta_{ij}= \int_{-\infty}^{\infty} {\rm d}x \,\phi_i(x) \phi_j(x),\quad
  \phi_{ijkl}= \int_{-\infty}^{\infty} {\rm d}x 
  \,\phi_i(x)\phi_j(x)\phi_k(x) \phi_l(x).
\end{equation}
Out of the sixteen combinations for $\phi_{ijkl}$, we only retain the
physically most relevant contributions and disregard others deliberately. This
leads to the following model Hamiltonian for two coupled Josephson junctions
\begin{eqnarray}
\label{eq:enden0}
  \hat{H}&=&  \Lambda(\eld\er+ \erd\el+\gld\gr+\grd\gl)\\\nonumber
  &+& \delta(\eld\el-\gld \gl)+ \delta(\erd\er-\grd \gr)\\
  &+&  \Omega_0(\eld \gl +\gld \el)+\Omega_1(\erd \gr +\grd \er)\\\nonumber
  &+&g( \eld \eld \el \el+\gld \gld \gl \gl+  
  \erd \erd \er \er+\grd \grd \gr \gr)
\end{eqnarray}
where all coupling constants are implicitly rescaled by the corresponding
single particle or two-body matrix elements and the new parameter $\Lambda$
measures the spatial hopping rate between the sites. 
\begin{figure}[h]
\begin{center}
  \includegraphics[width=0.5\columnwidth]{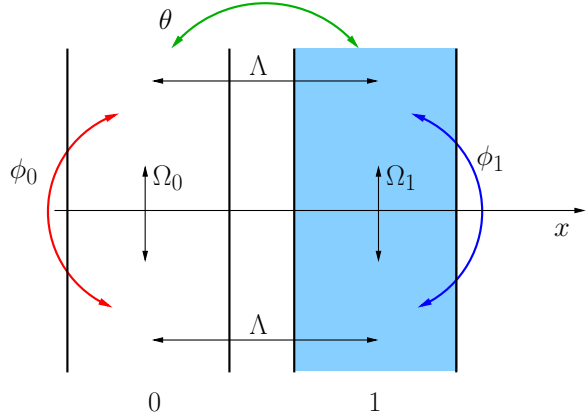}
\caption[Phases in two coupled Josephson junctions.]{
  Coupling pattern of two coupled Josephson junctions $JJ_0$ and $JJ_1$ with,
  $\Omega_1=-\Omega_0$.  The classical phase $\theta$ refers to a particle
  current between the spatial sites, while the phases $\phi_0$ and $\phi_1$
  are currents within the two-state atoms.}
\label{fig:modelangles}
\end{center}
\end{figure}

\subsection{Fock-space representation of the four mode model}
\begin{figure}[h]
\begin{center}
  \includegraphics[width=0.7\columnwidth]{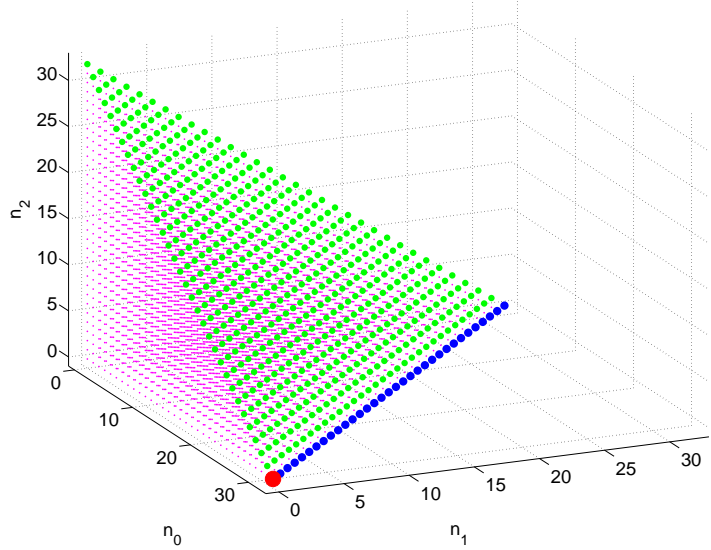}
\caption[4-d Fock space]{A three dimensional simplex represents the 
  four-mode Fock space for $N=32$ particles. The axes are labeled in a generic
  lexicographical order $(n_0,n_1,n_2)$ and implicitly $n_4=N-n_0-n_1-n_2$.}
\label{fock4d}
\end{center}
\end{figure}

In principle, it is possible to solve the four-mode Schr\"odinger equation in
Fock space by projecting it on the $N$-particle sector
\begin{equation}
  \label{eq:f4d}
  \ket{\psi,N}=  \delta(\hat{N}-N)
\sum_{\mathbf{n}=0}^\infty
\psi_{\mathbf{n}} \ket{\mathbf{n}=(n_0^e,n_0^g,n_1^e,n_1^g)}.
\end{equation}
The number constraint on the state  remains valid throughout the
time-evolution as number conservation is encoded into our Hamiltonian from the
beginning in Eq.~(\ref{numsym}).  This reduces the discrete $d=4$ dimensional
eigenvalue problem to an effective three-dimensional problem with nontrivial
boundaries. We have illustrated the finite support of the amplitude field
$\psi_{\mathbf{n}}$ in Fig.~\ref{fock4d}. It is a
$(d-1)$-dimensional simplex embedded into a $d$-dimensional Fock space.
The full analysis of this problem is an interesting problem in its own right
and will be presented in a forthcoming publication.

\subsection{The classical limit of the four mode model}
It is not necessary to solve the four-mode problem in Fock-space to understand
the principal features of the equilibrium configuration. Thus, we will again
resort to the classical approximation and use the number-symmetry broken
coherent state approximation for the quantum state
\begin{eqnarray}
  \label{eq:f4dcoh}
  \ket{\psi}&=&\ket{\boldsymbol{\alpha}=(e_0,g_0,e_1,g_1)^\top},\quad
  \hat{e}_{i}\ket{\psi}=e_i\ket{\psi},\quad
  \hat{g}_{i}\ket{\psi}=g_i\ket{\psi}.
\end{eqnarray}
The dynamics is simply obtained from the Lagrangian
\begin{eqnarray}
  \label{eq:lagcoh}
  \mathcal{L}(\boldsymbol{\alpha},\dot{\boldsymbol{\alpha}})&=&
  \bra{\boldsymbol{\alpha}}i\partial_t-
  \hat{H}\ket{\boldsymbol{\alpha}}
  =i\boldsymbol{\alpha}^\ast \dot{\boldsymbol{\alpha}}
  -\mathcal{H}(\boldsymbol{\alpha},\boldsymbol{\alpha}^\ast)
 -\frac{i}{2}
\frac{d}{dt}\mathcal{N}(\boldsymbol{\alpha},\boldsymbol{\alpha}^\ast),
\end{eqnarray}
if we introduce the classical Hamilton functions
$\mathcal{H}(\boldsymbol{\alpha},\boldsymbol{\alpha}^\ast)
=\bra{\boldsymbol{\alpha}}\hat{H}\ket{\boldsymbol{\alpha}}$ and number
expectation values $\mathcal{N}(\boldsymbol{\alpha},\boldsymbol{\alpha}^\ast)
=
\bra{\boldsymbol{\alpha}}\hat{N}\ket{\boldsymbol{\alpha}}$, as
\begin{eqnarray}
  \mathcal{H}(\boldsymbol{\alpha},\boldsymbol{\alpha}^\ast)
&=&\Lambda\left(e_0^\ast e_1+e_1^\ast e_0+
    g_0^\ast g_1+g_1^\ast g_0\right)\\\nonumber
  &+&\delta_0\left(|e_0|^2-|g_0|^2\right)+
  \delta_1 \left(|e_1|^2-|g_1|^2\right)\\\nonumber
  &+&\Omega_0\left(
   e_0^\ast g_0+g_0^\ast e_0\right)+
  \Omega_1\left(   e_1^\ast g_1+g_1^\ast e_1\right)\\\nonumber
  &+&g(|e_0|^4+|g_0|^4+|e_1|^4+|g_1|^4),\\
  \label{eq:num}
  \mathcal{N}(\boldsymbol{\alpha},\boldsymbol{\alpha}^\ast)
  &=&|e_0|^2+|g_0|^2+|e_1|^2+|g_1|^2.
\end{eqnarray}
If the dynamical coordinate is $\boldsymbol{\alpha}$, then we find the 
canonical momentum as 
\begin{equation}
\pi_k=\frac{\partial \mathcal{L}}{\partial \dot{\alpha}_k}=i\alpha_k^\ast,
\end{equation}
Consequently, variables and momenta obey the Poisson bracket
$\left\{\alpha_j,\pi_k\right\}=\delta_{jk}$. By construction, number
conservation is satisfied dynamically as
\begin{equation}
  \label{eq:number}
  \frac{d}{dt}\mathcal{N}=\left\{\mathcal{H},\mathcal{N}\right\}=0
\end{equation}
and the Hamilton equations of motion for the coordinates read
\begin{eqnarray}
\frac{d}{dt}\boldsymbol{\alpha}=\left\{\mathcal{H},\boldsymbol{\alpha}\right\}=-i
\mathcal{K} \boldsymbol{\alpha}\\
\mathcal{K}=
\left(
  \begin{array}{cccc}
    \delta_0+2g|\alpha_0|^2&\Omega_0&\Lambda&0\\
    \Omega_0&-\delta_0+2g|\alpha_1|^2&0&\Lambda\\
    \Lambda&0&\delta_1+2g|\alpha_2|^2&\Omega_1\\
    0&\Lambda&\Omega_1&-\delta_1+2g|\alpha_3|^2
  \end{array}
\right).\nonumber
\end{eqnarray}

In classical mechanics, 
we can deliberately choose new
coordinates and momenta to accounts for symmetries of the Hamiltonian. If a
new variable of a canonical transformation matches a conserved quantity, it
follows that the conjugate variable becomes cyclic.  In this spirit, we will
introduce the following pairs of action-angle variables: $(\Phi,\mathcal{N})$
measures the global phase and the total particle number
$\mathcal{N}=n_0^g+n_0^e+n_1^g+n_1^e$ of the system, $(\theta,\mathcal{M})$
measures the relative phase between left and right sites and the population
imbalance $\mathcal{M}=n_0^e+n_0^g-(n_1^e+n_1^g)$ inbetween,
$(\phi_0,\mathcal{M}_0=n_0^e-n_0^g)$, and $(\phi_1,\mathcal{M}_1=n_1^e-n_1^g)$
measure the relative internal phase of the atoms on each site and the
corresponding population difference.  This coupling scheme for the phases and
population imbalances has been illustrated in Fig.~\ref{fig:modelangles}.  By
inverting the population relations, one finds the individual occupation number
$n_l^\sigma$ per site as
\begin{eqnarray}
  \begin{array}{ll}
    n_0^e=\frac{1}{4}(\mathcal{N}+\mathcal{M}+2\mathcal{M}_0),&
    n_0^g=\frac{1}{4}(\mathcal{N}+\mathcal{M}-2\mathcal{M}_0),\\
    n_1^e=\frac{1}{4}(\mathcal{N}-\mathcal{M}+2\mathcal{M}_1),&
    n_1^g=\frac{1}{4}(\mathcal{N}-\mathcal{M}-2\mathcal{M}_1),
  \end{array}
\end{eqnarray}
Finally, we can use these physical coordinates in a canonical
transformation from complex amplitudes to real action-angle variables
\begin{eqnarray}
\label{acang}
  e_0&=& e^{-i(\Phi+\theta+\phi_0)}\sqrt{n_0^e}, 
  \quad g_0= e^{-i(\Phi+\theta-\phi_0)}\sqrt{n_0^g}\\\nonumber
  e_1&=& e^{-i(\Phi-\theta+\phi_1)}\sqrt{n_1^e}, 
  \quad g_1=e^{-i(\Phi-\theta-\phi_1)}\sqrt{n_1^g}.
\end{eqnarray}
For later use it is also useful to introduce the auxiliary phases $\theta_i$,
which are global phases of the subsystem on site i:
\begin{equation}
\label{globloc}
  \theta_0=\Phi+\theta,\quad \theta_1=\Phi-\theta.
\end{equation}
By substituting field amplitudes into the Lagrangian of Eq.~(\ref{eq:lagcoh}),
one can again identify the variables and corresponding canonical momenta as 
\begin{equation}
  \label{eq:mom}
    \frac{\partial \mathcal{L}}{\partial \dot{\Phi}}=\mathcal{N}, \quad
    \frac{\partial \mathcal{L}}{\partial \dot{\theta}}=\mathcal{M},\quad
    \frac{\partial \mathcal{L}}{\partial \dot{\phi}_0}=\mathcal{M}_0,\quad
    \frac{\partial \mathcal{L}}{\partial \dot{\phi}_1}=\mathcal{M}_1. 
\end{equation}
With this new coordinates and momenta we obtain a Hamilton function
\begin{eqnarray}
 \label{eq:HamilFuncFin}
 \mathcal{H}^\prime&=&\mathcal{H}^\prime
 (\theta,\phi_0,\phi_1,\mathcal{N},
 \mathcal{M},\mathcal{M}_0,\mathcal{M}_1)\\\nonumber
 &=&\frac{\Lambda}{2} \cos{(2\theta+\phi_0-\phi_1)}
 \sqrt{(\mathcal{N}+\mathcal{M}+2\mathcal{M}_0)
   (\mathcal{N}-\mathcal{M}+2\mathcal{M}_1)}\\\nonumber
 &+&\frac{\Lambda}{2}  \cos{(2\theta-\phi_0+\phi_1)}
 \sqrt{(\mathcal{N}+\mathcal{M}-2\mathcal{M}_0)
    (\mathcal{N}-\mathcal{M}-2\mathcal{M}_1)}\\\nonumber
  &+&\Omega_0 \cos{(2 \phi_0)}
  \sqrt{(\mathcal{N}+\mathcal{M})^2/4-\mathcal{M}_0^2}
  +\delta_0\mathcal{M}_0+\delta_1\mathcal{M}_1\\\nonumber
  & +&\Omega_1 \cos{(2\phi_1)}
  \sqrt{(\mathcal{N}-\mathcal{M})^2/4-\mathcal{M}_1^2}
  +  \frac{g}{4}\left(\mathcal{N}^2+\mathcal{M}^2
    +2\mathcal{M}_0^2+2\mathcal{M}_1^2\right).\nonumber
\end{eqnarray}

For the parameters $\delta=1$, $\Omega_0=1$, $\Omega_1=-1$, $\Lambda=0.1$,
$g=0.005$, $\mathcal{N}=100$, we have numerically minimized the energy of
Eq.~(\ref{eq:HamilFuncFin}) and find two minima $E=E_0,E_1$.  Please note that
the amplitudes of the solutions $\boldsymbol{\alpha}$ are real valued. Due to
the invariance of the solutions under global phase change, we can deliberately
modify them and choose the global phase of the left site as the common ground
$\theta_0=0$.  In Tab.~\ref{labelt}, we have listed the energies, the real
valued amplitudes and the phases calculated according Eqs.~(\ref{acang}) and
(\ref{globloc}).
\begin{table}[h]
\caption{\label{labelt} Two stationary solutions of the Hamilton equation,
  energies $E_0,E_1$, field amplitudes $\boldsymbol{\alpha}=(e_0,g_0,e_1,g_1)$ 
and phases. The global phase of the left site,
  \ie  
$\theta_0=0$
  was chosen as the common ground.}
\begin{indented}
\item[]
\begin{tabular}{@{}r|rrrr|rrrr|rr}
\br
$E$ & $e_0$& $g_0$ & $e_1$ & $g_1$ & $\Phi$ & $\phi_0$ & $\phi_1$ & $\theta$
&$\theta_0$&$\theta_1$\\
\mr
$-129.84$&  $2.82$ & $-6.48$ &   $2.82$  &  $6.48$& 
$\frac{\pi}{4}$ & $\frac{\pi}{2}$& 0&$-\frac{\pi}{4}$& 0&$\frac{\pi}{2}$\\
$-117.00$ &  $3.16$  & $-6.32$  & $-3.16$  & $-6.32$ & 
$-\frac{\pi}{4}$& $\frac{\pi}{2}$ &0 & $\frac{\pi}{4}$&0&$-\frac{\pi}{2}$\\
\br
\end{tabular}
\end{indented}
\end{table}
\begin{figure}[h]
\begin{center}
  \begin{tabular}{cc}
    \includegraphics[angle=0,width=0.5\columnwidth]{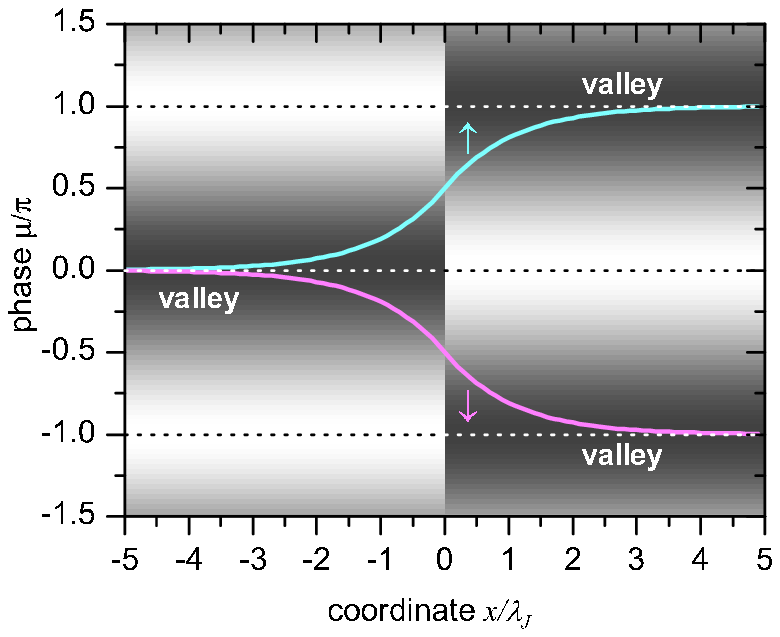}&
     \includegraphics[width=0.45\columnwidth]{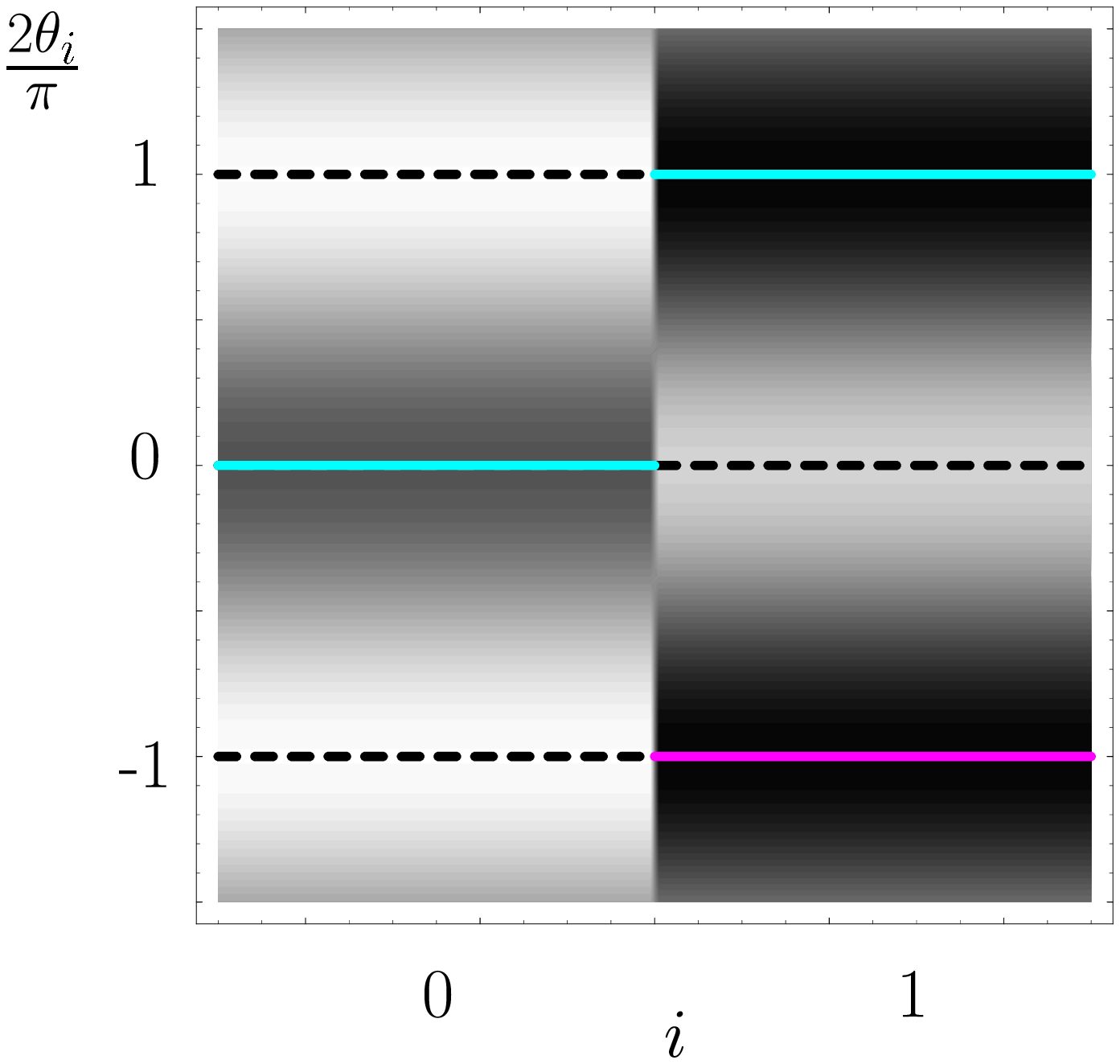}
  \end{tabular}
\caption[]{Comparing a spatially extended $0$-$\pi$-junction in a superconductor (left
  panel) calculated according to the sine-Gordon equation
  \cite{Goldobin:2005:MQC-2SFs} to an analog, two site junction in an atomic
  BEC (right panel). We have plotted the scaled global phase $\theta_i$ vs.
  the site index $i=0,1$.}
\end{center}
\end{figure}

\section{Conclusion and Perspectives}
\label{concl}
In the present article we have briefly summarized the status of the fluxon and
semifluxon physics in superconductivity. In particular, we were focusing on an
effect that exists in a long $0$-$\pi$ geometry where two Josephson junctions
were in the groundstate and one obtains a Josephson vortex of fractional
magnetic flux, pinned at the $0$-$\pi$ boundary.  Depending on the preparation
procedure, this ``classical'' state of the superconducting device can occur in
two different configurations: with the magnetic flux equal to $+\Phi_0/2$ or
$-\Phi_0/2$.

This arrangement of superconducting junctions has been analyzed and transfered
to the context of an atomic bosonic quantum gas, where two-state atoms in a
double well trap are coupled in a similar fashion.  There, the optical
$0$-$\pi$ junction is represented by the left/right localized internal atomic
Josephson junctions and a jumping phase of the complex-valued Rabi-frequency.
We have derived a simple four-mode model for this case and showed that in the
``classical'' approximation it qualitatively resembles the semifluxons seen in
superconductivity.

In the superconducting case, one observes a smooth spatial behaviour of the
phase across the $0$-$\pi$ boundary. In the atomic case, this will emerge also
by a more realistic modeling for the Rabi-frequency and spatial motion. This
is currently under investigation. Eventually, the four-mode, $0$-$\pi$
Josephson model will be also instrumental in examining the quantum properties
of these macroscopic semifluxon states and we will explore the macroscopic
tunneling between them \cite{Goldobin:2005:MQC-2SFs}.  This is also work in
progress and will be reported in a forthcoming publication.

\ack
The authors acknowledge the support of this work by the Deutsche
Forschungsgemeinschaft via the SFB/TRR 21, which is a collaboration of the
Universities of Stuttgart, T\"ubingen and Ulm, as well as the  
Max-Planck Institute for Solid-state Physics in Stuttgart, as well as
stimulating discussions with K. Vogel.


\section*{References}
\bibliographystyle{vancouver} 
\bibliography{bec,SF,jj-intrinsic,pi,SFS,MyPublications,ref_droptower}

\begin{thebibliography}{10}

\bibitem{enricofermiproc92}
Arimondo E, Phillips WD, Strumia F, editors.
\newblock Laser Manipulation of Atoms and Ions. North-Holland; 1992.

\bibitem{stringaribuch}
Pitaevskii L, Stringari S.
\newblock Bose-Einstein Condensation.
\newblock Oxford: Claredon Press; 2003.

\bibitem{kleinerbuch}
Buckel W, Kleiner R.
\newblock Superconductivity: Fundamentals and Applications.
\newblock 2nd ed. Wiley-VCH; 2004.

\bibitem{folman02}
Folman R, Kr{\"u}ger P, Schmiedmayer J, Denschlag J, Henkel C.
\newblock Mircoscopic atom optics: From wires to an atom chip.
\newblock Adv At Mol Opt Phys. 2002;48:263.

\bibitem{fortagh07}
Fort{\'a}gh J, Zimmermann C.
\newblock Magnetic microtraps for ultracold atoms.
\newblock Rev Mod Phys. 2007;79:235.
\newblock And Refs. therein.

\bibitem{anderson04}
Du S, Squires M, Imai Y, Czaia L, Saravanan R, Bright V, et~al.
\newblock Atom-chip Bose-Einstein condensation in a portable vacuum cell.
\newblock Phys~Rev~A. 2004;70:53606.

\bibitem{becmugrav06}
Vogel A, Schmidt M, Sengstock K, Bongs K, Lewoczko W, Schuldt T, et~al.
\newblock Bose-Einstein condensates in microgravity.
\newblock Appl Phys B. 2006;84:664.

\bibitem{nandi06}
Nandi G, Walser R, Kajari E, Schleich WP.
\newblock Dropping cold quantum gases on Earth over long times and large
  distances.
\newblock Phys~Rev~A. 2007;76:63617.

\bibitem{koenemann07}
K{\"o}nemann T, Brinkmann W, G{\"o}kl{\"u} E, L{\"a}mmerzahl C, Dittus H, van
  Zoest T, et~al.
\newblock First Realization of a magneto-optical trap in weightlessness.
\newblock Appl Phys B. 2007;89:431.

\bibitem{likharev79}
Likharev K.
\newblock Superconducting weak links.
\newblock Rev Mod Phys. 1979;51:101.

\bibitem{barone82}
Barone A, Paterno G.
\newblock Physics and Application of the Josephson Effect.
\newblock New York: Wiley Interscience; 1982.

\bibitem{schoen01}
Makhlin Y, Sch{\"o}n G, Shnirman A.
\newblock Quantum-state engineering with Josephson-junction devices.
\newblock Rev Mod Phys. 2001;73:357.

\bibitem{castin97}
Castin Y, Dalibard J.
\newblock Relative phase of two Bose-Einstein condensates.
\newblock Phys~Rev~A. 1997;55:4330.

\bibitem{ketterleinterference}
Andrews M, Townsend C, Miesner HJ, Durfee D, Kurn D, Ketterle W.
\newblock Observation of Interference Between Two Bose Condensates.
\newblock Science. 1997;275:637.

\bibitem{ketterleProc99}
Ketterle W, Durfee D, Stamper-Kurn D.
\newblock Making, probing and understanding Bose-Einstein condensates.
\newblock In: Inguscio M, Stringari S, Wieman C, editors. Proceedings of the
  International School of Physics "Enrico Fermi", Course CXL. Soc. Italiana di
  Fisica,Bologna, Italy. Amsterdam: IOS Press; 1999. .

\bibitem{Anderson1998a}
Anderson BP, Kasevich MA.
\newblock Macroscopic Quantum Interference from Atomic Tunnel Arrays.
\newblock Science. 1998 Nov;282:1686.

\bibitem{sols98}
Zapata I, Sols F, Leggett T.
\newblock Josephson effect between trapped Bose-Einstein condensates.
\newblock Phys~Rev~A. 1998;57:R28.

\bibitem{williams199}
Williams J, Walser R, Cooper J, Cornell E, Holland M.
\newblock Nonlinear \protect{Josephson-type} oscillations of a driven,
  \protect{two-component} \protect{Bose-Einstein} condensate.
\newblock Phys Rev A, Rapid Comm. 1999;59:31.

\bibitem{williams499}
Williams J, Walser R, Cooper J, Cornell E, Holland M.
\newblock Excitation of a dipole topological state in a strongly coupled
  two-component Bose-Einstein condensate.
\newblock Phys~Rev~A. 2000;61:033612.

\bibitem{leggett401}
Leggett A.
\newblock Bose-Einstein condensation in the alkali gases: Some fundamental
  concepts.
\newblock Rev Mod Phys. 2001;73:307.

\bibitem{giovanazzi500}
Giovanazzi S, Smerzi A, Fantoni S.
\newblock Josephson Effects in Dilute Bose-Einstein Condensates.
\newblock Phys~Rev~Lett. 2000;84:4521.

\bibitem{oberthaler05}
Albiez M, Gati R, F{\"o}lling J, Hunsmann S, Cristian M, Oberthaler MK.
\newblock Direct observation of tunneling and nonlinear self-trapping in a
  single bosonic Josephson junction.
\newblock Phys~Rev~Lett. 2005;95:010402.

\bibitem{oberthaler07}
Gati R, Oberthaler M.
\newblock A bosonic Josephson junction.
\newblock J Phys B: At Mol Opt Phys. 2007;40:R61.

\bibitem{inguscio801}
Cataliotti F, Burger S, Fort C, Maddaloni P, Minardi F, Trombettoni A, et~al.
\newblock Josephson junction arrays with Bose-Einstein condensates.
\newblock Science. 2001;293:843.

\bibitem{nandi07}
Nandi G, Sizman A, Fort\'a{}gh J, Walser R.
\newblock A numberfilter for matterwaves.
\newblock arXiv: 07101737. 2007;.

\bibitem{bloch05}
Bloch I.
\newblock Ultracold quantum gases in optical lattices.
\newblock Nature Physics. 2005;1:23.

\bibitem{bloch07}
F{\"o}lling S, Trotzky S, Cheinet P, Feld M, Saers R, Widera1 A, et~al.
\newblock Direct observation of second-order atom tunnelling.
\newblock Nature. 2007;448:1029.

\bibitem{dalibard04}
Hadzibabic Z, Stock S, Battelier B, Bretin V, Dalibard J.
\newblock Interference of an array of independent Bose-Einstein condensates.
\newblock Phys~Rev~Lett. 2004;93:180403.

\bibitem{kaurov:013627}
Kaurov VM, Kuklov AB.
\newblock Atomic Josephson vortices.
\newblock Phys~Rev~A. 2006;73(1):013627.

\bibitem{kaurov:011601}
Kaurov VM, Kuklov AB.
\newblock Josephson vortex between two atomic Bose-Einstein condensates.
\newblock Phys~Rev~A. 2005;71(1):011601.

\bibitem{Goldobin:SF-Shape}
Goldobin E, Koelle D, Kleiner R.
\newblock {Semifluxons in long {J}osephson 0-$\pi$-junctions}.
\newblock Phys Rev B. 2002;66:100508(R).

\bibitem{Kleiner:1992:IJE}
Kleiner R, Steinmeyer F, Kunkel G, M\"uller P.
\newblock Intrinsic {J}osephson effects in Bi$_2$Sr$_2$CaCu$_2$O$_8$ single
  crystals.
\newblock Phys Rev Lett. 1992 Apr;68(15):2394--2397.

\bibitem{Kleiner:1994:IJE}
Kleiner R, M\"uller P.
\newblock Intrinsic {J}osephson effects in high-$T_{c}$ superconductors.
\newblock Phys Rev B. 1994 Jan;49(2):1327--1341.

\bibitem{Ryazanov:2001:SFS-PiJJ}
Ryazanov VV, Oboznov VA, Rusanov AY, Veretennikov AV, Golubov AA, Aarts J.
\newblock Coupling of Two Superconductors through a Ferromagnet: Evidence for a
  $\pi$ Junction.
\newblock Phys Rev Lett. 2001;86:2427.

\bibitem{Blum:2002:IcOscillations}
Blum Y, Tsukernik A, Karpovski M, Palevski A.
\newblock {Oscillations of the Superconducting Critical Current in
  Nb-Cu-Ni-Cu-Nb Junctions}.
\newblock Phys Rev Lett. 2002;89:187004.

\bibitem{Bauer:2004:SFS-SpontSuperCurrents}
Bauer A, Bentner J, Aprili M, Della-Rocca ML, Reinwald M, Wegscheider W, et~al.
\newblock Spontaneous Supercurrent Induced by Ferromagnetic $\pi$ Junctions.
\newblock Phys Rev Lett. 2004;92(21):217001.

\bibitem{Sellier:2004:SFS:HalfIntShapiro}
Sellier H, Baraduc C, Lefloch F, Calemczuk R.
\newblock Half-Integer Shapiro Steps at the $0$-$\pi$ Crossover of a
  Ferromagnetic {J}osephson Junction.
\newblock Phys Rev Lett. 2004;92(25):257005.

\bibitem{Oboznov:2006:SFS-Ic(dF)}
Oboznov VA, Bol'ginov VV, Feofanov AK, Ryazanov VV, Buzdin AI.
\newblock Thickness Dependence of the {J}osephson Ground States of
  Superconductor-Ferromagnet-Superconductor Junctions.
\newblock Phys Rev Lett. 2006;96(19):197003.

\bibitem{Kontos:2002:SIFS-PiJJ}
Kontos T, Aprili M, Lesueur J, Gen\^et F, Stephanidis B, Boursier R.
\newblock {J}osephson Junction through a Thin Ferromagnetic Layer: Negative
  Coupling.
\newblock Phys Rev Lett. 2002;89:137007.

\bibitem{Weides:2006:SIFS-HiJcPiJJ}
Weides M, Kemmler M, Goldobin E, Koelle D, Kleiner R, Kohlstedt H, et~al.
\newblock High quality ferromagnetic 0 and $\pi$ {J}osephson tunnel junctions.
\newblock Appl Phys Lett. 2006;89(12):122511.

\bibitem{Vavra:2006:SIfIS}
Vavra O, Gazi S, Golubovic DS, Vavra I, Derer J, Verbeeck J, et~al.
\newblock 0 and $\pi$ phase {J}osephson coupling through an insulating barrier
  with magnetic impurities.
\newblock Phys Rev B. 2006;74(2):020502.

\bibitem{Cleuziou:2006:CNT-SQUID}
Cleuziou JP, Wernsdorfer W, Bouchiat V, Ondarcuhu T, Monthioux M.
\newblock Carbon nanotube superconducting quantum interference device.
\newblock Nature Nanotech. 2006;1(1):53--59.

\bibitem{vanDam:2006:QuDot:SuperCurrRev}
van Dam JA, Nazarov YV, Bakkers EPAM, De~Franceschi S, Kouwenhoven LP.
\newblock Supercurrent reversal in quantum dots.
\newblock Nature (London). 2006;442(7103):667--670.

\bibitem{Bulaevskii:pi-loop}
Bulaevski\u{i} LN, Kuzi\u{i} VV, Sobyanin AA.
\newblock Superconducting system with weak coupling to the current in the
  ground state.
\newblock JETP Lett. 1977;25(7):290--294.
\newblock [Pis'ma Zh. Eksp. Teor. Fiz. 25, 314 (1977)].

\bibitem{Bulaevskii:0-pi-LJJ}
Bulaevskii LN, Kuzii VV, Sobyanin AA.
\newblock On Possibility of the Spontaneous Magnetic Flux in a {J}osephson
  junction containing magnetic impurities.
\newblock Solid State Commun. 1978;25:1053--1057.

\bibitem{Xu:SF-shape}
Xu JH, Miller JH, Ting CS.
\newblock $\pi$-vortex state in a long 0-$\pi$-{J}osephson junction.
\newblock Phys Rev B. 1995;51:11958.

\bibitem{Kirtley:SF:T-dep}
Kirtley JR, Tsuei CC, Moler KA.
\newblock Temperature Dependence of The Half-Integer Magnetic Flux quantum.
\newblock Science. 1999;285:1373.

\bibitem{Kirtley:SF:HTSGB}
Kirtley JR, Tsuei CC, Rupp M, Sun JZ, Yu-Jahnes LS, Gupta A, et~al.
\newblock Direct imaging of integer and half-integer {J}osephson vortices in
  high-{$T_c$} grain boundaries.
\newblock Phys Rev Lett. 1996;76:1336.

\bibitem{Smilde:ZigzagPRL}
Smilde HJH, Ariando, Blank DHA, Gerritsma GJ, Hilgenkamp H, Rogalla H.
\newblock $d$-Wave-Induced {J}osephson Current Counterflow in
  {YBa$_2$Cu$_3$O$_7$/Nb} Zigzag Junctions.
\newblock Phys Rev Lett. 2002;88:057004.

\bibitem{Ariando:Zigzag:NCCO}
Ariando, Darminto D, Smilde HJH, Leca V, Blank DHA, Rogalla H, et~al.
\newblock Phase-Sensitive Order Parameter Symmetry Test Experiments Utilizing
  {Nd$_{2-x}$Ce$_x$CuO$_{4-y}$/Nb} Zigzag Junctions.
\newblock Phys Rev Lett. 2005;94(16):167001.

\bibitem{Hilgenkamp:zigzag:SF}
Hilgenkamp H, Ariando, Smilde HJH, Blank DHA, Rijnders G, Rogalla H, et~al.
\newblock Ordering and manipulation of the magnetic moments in large-scale
  superconducting $\pi$-loop arrays.
\newblock Nature (London). 2003;422:50--53.

\bibitem{Kirtley:2005:AFM-SF}
Kirtley JR, Tsuei CC, Ariando, Smilde HJH, Hilgenkamp H.
\newblock Antiferromagnetic ordering in arrays of superconducting $\pi$-rings.
\newblock Phys Rev B. 2005;72(21):214521.

\bibitem{Kogan:3CrystalVortices}
Kogan VG, Clem JR, Kirtley JR.
\newblock {J}osephson vortices at tricrystal boundaries.
\newblock Phys Rev B. 2000;61(13):9122--9129.

\bibitem{Zenchuk:2003:AnalXover}
Zenchuk A, Goldobin E.
\newblock Analytical analysis of ground states of 0-$\pi$ long {J}osephson
  junctions.
\newblock Phys Rev B. 2004;69:024515.

\bibitem{Susanto:SF-gamma_c}
Susanto H, van Gils SA, Visser TPP, Ariando, Smilde HJH, Hilgenkamp H.
\newblock Static semifluxons in a long {J}osephson junction with
  $\pi$-discontinuity points.
\newblock Phys Rev B. 2003;68:104501.

\bibitem{Kirtley:IcH-PiLJJ}
Kirtley JR, Moler KA, Scalapino DJ.
\newblock {Spontaneous flux and magnetic-interference patterns in 0-$\pi$
  {J}osephson junctions}.
\newblock Phys Rev B. 1997;56:886.

\bibitem{Goldobin:F-SF}
Goldobin E, Stefanakis N, Koelle D, Kleiner R.
\newblock Fluxon-semifluxon interaction in an annular long {J}osephson
  $0$-$\pi$ junction.
\newblock Phys Rev B. 2004;70(9):094520.

\bibitem{Lazarides:Ic(H):SF-Gen}
Lazarides N.
\newblock Critical current and fluxon dynamics in overdamped 0-$\pi$
  {J}osephson junctions.
\newblock Phys Rev B. 2004;69(21):212501.

\bibitem{Goldobin:SF-ReArrange}
Goldobin E, Koelle D, Kleiner R.
\newblock Ground state and bias current induced rearrangement of semifluxons in
  0-$\pi$-{J}osephson junctions.
\newblock Phys Rev B. 2003;67:224515.

\bibitem{Susanto:2005:1D-FractVortexCrystal}
Susanto H, Goldobin E, Koelle D, Kleiner R, van Gils SA.
\newblock Controllable plasma energy bands in a one-dimensional crystal of
  fractional {J}osephson vortices.
\newblock Phys Rev B. 2005;71(17):174510.

\bibitem{Goldobin:2005:MQC-2SFs}
Goldobin E, Vogel K, Crasser O, Walser R, Schleich WP, Koelle D, et~al.
\newblock Quantum tunneling of semifluxons in a $0$-$\pi$-$0$ long {J}osephson
  junction.
\newblock Phys Rev B. 2005;72(5):054527.

\bibitem{schleich01}
Schleich WP.
\newblock Quantum Optics in Phase Space.
\newblock Berlin, Germany: Wiley-VCH; 2001.

\bibitem{feynman}
Feynman RP, Hibbs AH.
\newblock Quantum mechanics and path integrals.
\newblock McGraw-Hill; 1965.

\end{thebibliography}

\end{document}